\documentstyle[epsf]{lamuphys}

\def\kms{\mbox{\rm km\,s$^{-1}$}}
\def\kpc{\mbox{\rm kpc}}
\def\etal{\mbox{\rm et~al.}}

\def\bh{{\sc bh}}

\begin{document}

\title{M87 as a Galaxy}
\author{Walter Dehnen}
\institute{Theoretical Physics, 1 Keble Road, Oxford OX1 3NP, United Kingdom}
\maketitle

\begin{abstract}

I review recent studies about the gravitational potential and stellar dynamics
of M87 in particular, and the dynamics of the stars in the presence of a 
super-massive central black hole, in general. 

At large radii, investigations of both the X-ray emitting gas and the velocity 
distribution of globular clusters indicate the presence of large amounts of 
non-luminous matter, possibly belonging to the inner parts of the Virgo cluster.

At small radii, there is no evidence from the stellar kinematics,
at most a hint, for the existence of a central point mass, whereas the gas 
dynamics reveal the presence of a highly concentrated mass in the centre of
M87, possibly a super-massive black hole (\bh). Given the existence of such 
a central mass, the stellar kinematics indicate a strong tangential anisotropy
of the stellar motion inside a few arcseconds. The implications of this result
for the evolution and formation history of M87 and its central \bh\ are 
discussed.
I also discuss in more general terms the structural changes that a highly 
concentrated central mass can induce in its parent galaxy.

\end{abstract}

\section{Introduction}
\index{galaxy formation}
According to their observable properties, elliptical galaxies can be divided 
into two classes. This dichotomy is most clearly revealed in the central 
brightness distribution (cf.\ Fig.~3 of Gebhardt \etal\ 1996). Consequently, 
the two classes are commonly called `power-law galaxies' and `core galaxies' 
(though -- as so often in astronomy -- the latter term is highly misleading: 
these galaxies actually do not have a core of constant density). Core galaxies 
have shallow central luminosity density profiles with $j\propto r^{-\gamma}$, 
$\gamma\la1.3$, ellipticities E0 to E3-4, elliptic to boxy isophotes, and 
negligible rotation $v_{\rm rot}\ll \sigma$. They are on average bright 
($M_V\la-19.5$) and often radio-loud and X-ray-active, possess extended stellar
envelopes and rich ($N\ga2000$), extended globular cluster (GC) systems 
which are multi-modal in their properties. These galaxies are thought to be 
of round to triaxial shape supported by anisotropy in the stellar motions.

Power-law galaxies, on the other hand, have steep central density cusps with 
$\gamma\ga-1.5$, ellipticities up to E7, elliptic to disky isophotes, and 
significant rotation velocities $v_{\rm rot}\sim\sigma$. They are on average 
fainter ($M_V\ga-21.5$) and show no radio or X-ray activity; their surface 
density follows a de Vaucouleurs profile and their GC systems are poor 
($N\la1500$) and with a profile following that of the stellar light. These 
galaxies are believed to be of near-oblate shape supported by rotational 
motions.

Clearly, M87 having a shallow density cusp ($\gamma\approx1.2$), round 
isophotes, neglible rotation, radio and X-ray activity, an extended stellar 
envelope, and a rich and bi-modal globular cluster system is a generic 
representative of the class of core galaxies. It is generally believed, that 
the core galaxies are formed by one or more major merger events. In face of 
this hypothesis, it is important to ask whether M87 is consistent with being a 
merger remnant. 

\begin{figure}
	\centerline{ \epsfxsize=8cm \epsfbox{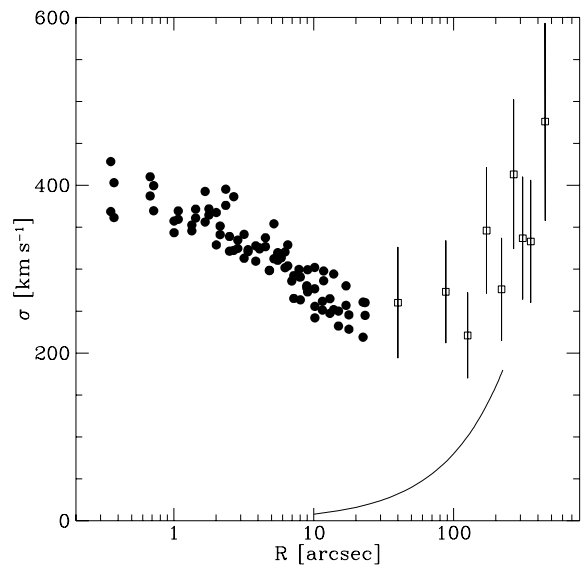}}
\caption[]{Velocity dispersion for M87 as measured for stars (filled circles, 
van der Marel 1994) and planetary nebul\ae\ (open squares, Cohen \& Ryzhov 
1997). The full line is the rotational velocity measured by the latter authors.}
	\label{fig-sigma}
\end{figure} 

\begin{figure}
	\centerline{ \epsfxsize=8cm \epsfbox{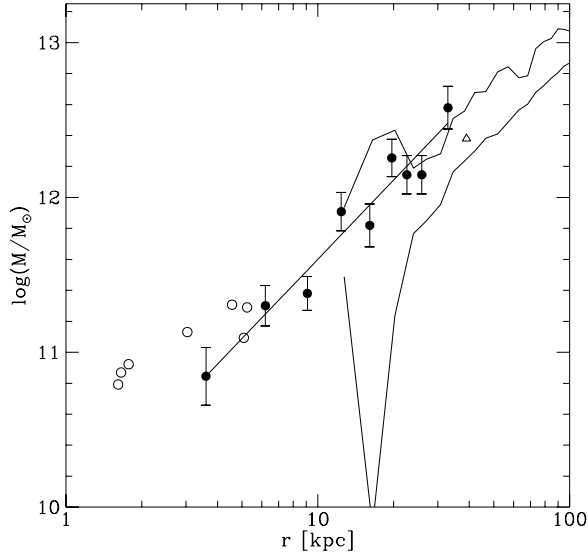}}
\caption[]{Mass enclosed in radius $r$ as derived from studies of stellar 
kinematics (open circles, Sargent \etal\ 1978), globular clusters (triangle, 
Mould \etal\ 1990, and filled circles, Cohen \& Ryzhov 1997), and X-ray gas 
(thin lines: lower and upper limits, Nulsen \& B\"ohringer 1995, corrected for
the difference in the adopted distance to M87). The solid line represents a 
power-law fit to the filled circles: ${\rm M}\propto r^{1.7}$.} 
\label{fig-mass}
\end{figure} 

\section{The Matter Distribution at Large Radii}
\index{globular clusters}
\index{globular clusters!kinematics}
\index{dark halo}
\index{mass of M87}
\index{rotation velocity}
\index{velocity dispersion}
The presence of dark matter around spiral galaxies was established by studies
of the motions of their gaseous disk, whose emission lines can be traced to
large radii. Elliptical galaxies in general have very little cold gas and one 
has traditionally used stars as dynamical tracer population. However, at large 
radii the stellar absorption-line spectra are very hard, if not impossible, to 
observe with useful signal-to-noise ratio. Thus, in order to probe the 
potential at large radii of elliptical galaxies, one needs different kinematical
tracers. Possible candidates are gas rings, planetary nebul\ae\ (PN), and 
globular clusters (GC). Gas rings are rather rare among elliptical galaxies, 
and in general one is left with PNs or GCs. The problem with using these or 
stars as tracers is that they do not move on circular orbits like the gas, but 
form dynamically hot systems, which complicates the interpretation of the 
measured kinematics.

Recently, Cohen \& Ryzhov (1997) have studied the GC system of M87. The
derived rotation velocity $v_{\rm rot}$ and velocity dispersion $\sigma$ are 
displayed in Fig.~\ref{fig-sigma} together with stellar kinematical data for 
the central parts. There is a clear change in the kinematic properties 
at about 100\arcsec\ from the centre: in the inner parts $v_{\rm rot}\approx 0$ 
and $\sigma$ decreases from $\approx400\kms$ to $\approx200\kms$; in the outer 
parts $\sigma$ increases reaching $\approx400\kms$ at 400\arcsec\ 
and $v_{\rm rot}$ becomes significant. This kinematical behaviour is very 
similar to that observed for NGC 1399 (cD galaxy in the Fornax cluster)
by Arnaboldi \& Freeman (1996) using PNs as tracers (their Fig.~\ref{fig-mass}).
An obvious explanation is that the change in kinematical properties constitutes
the transition from the highly concentrated galaxy to the less concentrated and 
dark-matter-dominated Virgo cluster.

Cohen \& Ryzhov have analysed their data in terms of the mass distribution
and found that in the range $3\kpc\le r\le30\kpc$ (using a distance to M87 of 
15\,Mpc) their data imply a mass density $\rho\propto r^{-1.3}$ and a mass of 
$\approx3 \times10^{12}{\rm M}_\odot$ inside 44\kpc. These findings are in 
agreement with estimates derived from X-ray observations (\cite{XrayA,XrayB}),
see Fig.~2. It is intriguing that the derived density slope of $\gamma=1.3$ 
agrees well with $1.4$ recently predicted for the inner parts of dark matter 
halos by high-resolution simulations of CDM cosmogonies (\cite{moor:etal:97}).

\section{Small Radii: Stellar Kinematics and the Black Hole}
\index{black hole}
\index{density cusp}
\index{kinematics!stellar}
\index{dynamics!stellar}
Because of its stellar kinematics, M87 has for a long time been suspected to 
harbour at its centre a super-massive black hole\footnote{
	Clearly, from stellar dynamical and similar arguments, one cannot
	infer that the dark object is of the size of its Schwarzschild radius.
	However, lacking other plausible explanations for the high mass 
	concentrations at the centres of many galaxies, it is generally 
	believed that these are actually black holes, and I adopt this 
	hypothesis throughout this paper.} (\bh), 
though this conclusion was always controversial (cf.\ \cite{sarg:etal:78,%
binn:mamm:82,merr:87}). The best currently available photometry 
(\cite{laue:etal:92}) and stellar kinematical data (\cite{vdM:94}) show a 
weak density cusp and a slightly centrally peaked velocity dispersion $\sigma$
(see Fig.~\ref{fig-sigma}). As van der Marel's analysis showed, these data can 
easily be interpreted by a pure stellar model with constant mass-to-light 
ratio and anisotropy%
\footnote{The anisotropy parameter is defined to be $$\beta\equiv1-\sigma_t^2/
	\sigma_r^2,$$ where $\sigma_t$ and $\sigma_r$ denote the tangential
	and radial velocity dispersions. For an isotropic system $\beta=0$;
	for radial anisotropy $\beta>0$ reaching $\beta=1$ for pure radial 
	orbits; and for tangential anisotropy $\beta<0$ with $\beta=-\infty$
	for pure circular orbits.}
of $\beta\approx0.5$ (\cite{vdM:94}), i.e.\ radially anisotropic as expected 
for a galaxy formed by violent relaxation. Alternatively, one can explain the
centrally peaked $\sigma$ by a central \bh\ of a few billion solar masses.
However, such a model does not work with radial anisotropy near the \bh, which
would give too high a central $\sigma$, and requires the opposite: isotropy
or tangential anisotropy, depending on the mass of the \bh.

Bender \etal\ (1994) found a change in the profile of the stellar absorption 
lines: the coefficient $h_4$ of the Gauss-Hermite expansion of the profile 
changes sign at $r\approx3\arcsec$, such that $h_4<0$ inside and $h_4>0$
outside. This implies a change in the underlying dynamical properties in the 
sense that $\beta_{r<3^{\prime\prime}}<\beta_{r>3^{\prime\prime}}$, as predicted
by models with central \bh\ (though it would be very hard to quantify this).

\begin{figure}
	\centerline{ \epsfxsize=8cm \epsfbox{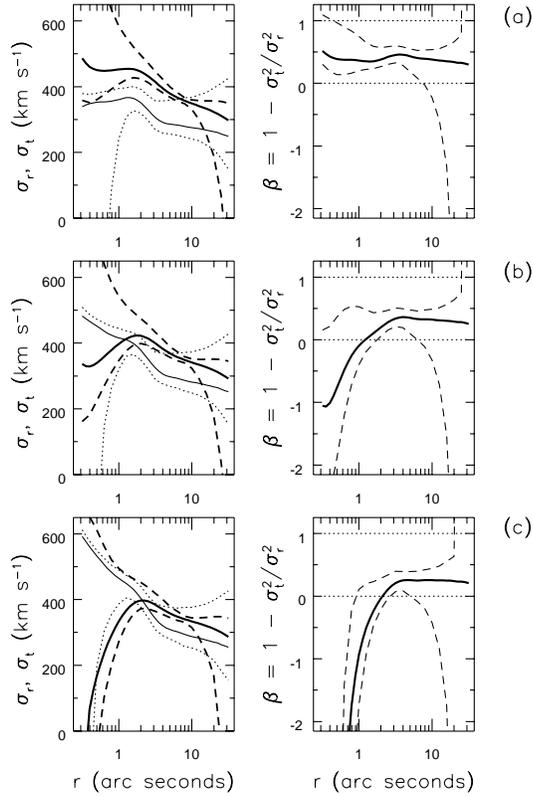}}
\caption[]{Velocity dispersions and anisotropy vs.\ radius for M87 as inferred
by Merritt \& Oh (1997) from the stellar kinematics for a \bh\ with assumed
mass of (a) 1, (b) 2.4, and (c) 3.8 billion solar masses. Dotted lines give 
95\% confidence bands.}\label{fig-oh}
\end{figure}

Actually, since 1994 we {\em know} from HST observations of a central gas-disk 
that M87 hosts a central \bh\ of ${\rm M}_\bullet=(3.2\pm0.9)\times10^9{\rm 
M}_\odot$ (\cite{harm:etal:94,ford:etal:94,marc:etal:97} and this volume). 
Starting from this fact and assuming spherical symmetry and a constant stellar 
mass-to-light ratio, Merritt \& Oh (1997) solved simultaneously the projection 
equation for the stellar kinematics and the Jeans equation describing dynamical 
equilibrium. This gave them a non-parametric estimate for the anisotropy implied
by the observed stellar kinematics and the \bh. Their results for a \bh\ of
1, 2.4, and 3.8 billion solar masses are displayed in Fig.~\ref{fig-oh}. For a
sufficiently massive \bh, the central stellar motions are strongly tangentially
anisotropic ($\beta\la-1$). The more massive the \bh, the stronger the inferred
anisotropy and the larger is the radius at which $\beta$ changes sign. Outside 
of this radius, the stellar motions are slightly radially anisotropic in 
agreement with van der Marel's (1994) study. For the most recent estimate of 
${\rm M} _\bullet$, the effect is very significant leading to highly tangential 
anisotropy. In the following section, I discuss possible explanations for this 
effect.

\section{Stellar Dynamics Around a Black Hole}
\index{black hole!formation}
\index{black hole!interaction with stars}
\index{density cusp!formation}
Can we understand the abrupt change in the stellar motions from radial to 
tangential anisotropy near the \bh? Let us consider various possible
scenarios for the formation of the \bh\ at the centre of a galaxy.

\bigskip
\noindent{\bf Adiabatic Growth of the Black Hole}\par\nobreak\noindent
If the \bh\ grows slowly over a long period of time, the actions of the 
stellar orbits are conserved. However, the mapping between actions and ordinary
phase-space coordinates changes as the potential evolves from harmonic (for 
an initially isothermal core) to Keplerian. This change results in a stellar 
density cusp with $\rho\propto r^{-3/2}$ and a mildly tangential anisotropy 
of at most $\beta=-0.3$ (\cite{youn:80,good:binn:84,quin:etal:95}) inside the 
sphere of influence of the \bh, which has radius
\begin{equation}
	r_h = {G\,{\rm M}_\bullet\over\sigma^2} 
	\big(\simeq 1\arcsec \mbox{ for M87}\big).
\end{equation}
While this radius is of the correct size, the effect is much too weak in order 
to explain Merritt \& Oh's result of $\beta\la-1$. Furthermore, M87 has a 
central density slope $\gamma\approx1.2<3/2$.

\bigskip
\noindent{\bf Growth by the Capture of Stars}\par\nobreak\noindent
A \bh\ may grow by tidally disrupting and capturing stars that happen to come
too close to it. Such stars will predominantly have low angular momenta, while
near-circular orbits are hardly affected. Hence, this process gives rise to
$\beta<0$. However, the distance from the \bh\ that a star must reach before it 
gets destroyed is of the order of the \bh's Schwarzschild radius, which is 
smaller than $r_h$ by several orders of magnitude.

\bigskip
\index{black hole!binary}
\noindent{\bf Growth by the Accretion of Black Holes}\par\nobreak\noindent
If two galaxies, each hosting a massive \bh\ at its centre, merge, the \bh s 
will sink to the centre of the remnant and finally merge as well 
(\cite{ebis:etal:91}). Begelman \etal\ (1980) have outlined the stages of this
process:
\begin{itemize}
\item[(1)]Due to dynamical friction with the background stars, the \bh s sink
	into the centre and form a \bh\ binary.
\item[(2)]The binary looses energy and angular momentum (it ``hardens'')
	due to three-body interactions with passing stars.
\item[(3)]When the separation between the \bh s has become sufficiently small, 
	gravitational radiation becomes very efficient in hardening the 
	binary until it finally merges to a single \bh.
\end{itemize}
Depending on the time-scale of the whole process, repeated events of this kind
are possible\footnote{
	One \bh-merger has to be finished before a third \bh\ arrives, since
	more than two \bh s cannot co-exist for long in a galaxy's centre, as 
	all but two of them will quickly be ejected by sling shots.},
in particular for a central-cluster elliptical, such as M87, which over its 
lifetime has likely cannabalised many minor companions. The least understood 
mechanism here is (2), which is also the process most relevant to the possible 
effect on the stellar dynamics. One problem, for instance, is whether or not the
eccentricity of the binary \bh\ increases, which in turn is relevant for (3), 
since for a highly eccentric binary gravitational radiation can take over at 
higher energies, i.e.\ earlier, than for a less eccentric binary. Another 
problem is loss-cone depletion: three-body interactions that harden the binary 
eject stars out of the centre, and the number of interaction candidates 
diminishes. Eventually, this may even halt the hardening before gravitational 
radiation can take over (the re-fueling of the loss cone due to two-body 
relaxation is much too slow).

It is clear that process (2) will create tangential anisotropy, since, as for 
the tidal disruption, low-angular-momentum stars are more likely to interact 
with the \bh s -- they are, however, not eaten by the \bh\ but ejected via a
sling-shot. Evidently, this process reaches out to the radius $r_h$, where the 
circular motion around the \bh s equals the mean stellar velocity dispersion.

\begin{figure}
	\centerline{ \epsfxsize=10cm \epsfbox{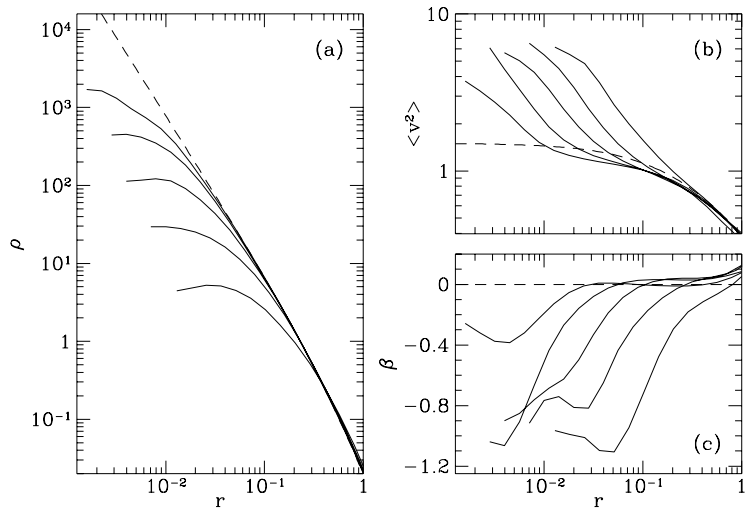}}
\caption[]{Response of a Jaffe model to the hardening of a \bh\ binary as 
computed by Quinlan \& Hernquist (1997). The dashed lines show the initial
(a) density, (b) velocity dispersion, and (c) anistotropy. The solid lines
show these quantities when the binary reaches a certain hardness. The binary's
components have equal mass of 0.04, 0.02, 0.01, 0.005, and 0.0025 (leftmost) 
in units of the total mass in stars.
{\scriptsize Reprint from New Astronomy, Vol.~2, Quinlan \& Hernquist, "The 
dynamical evolution of massive black hole binaries - II. Self-consistent 
$N$-body integrations", p.~533-554, 1997 with kind permission of Elsvier 
Science - NL, Sara Burgerhartstraat 25, 1055 KV Amsterdam, The Netherlands.}
}\label{quinlan_hernquist}
\end{figure}

Recently, Quinlan (1996) and Quinlan \& Hernquist (1997) have studied processes
(1) and (2) by detailed simulations. They developed an $N$-body code in which
the \bh-\bh\ interaction is computed exactly (Newtonianly; relativistic effects 
are unimportant before stage 3), the \bh-star interactions by a softened 
Keplerian force, and the star-star interactions via an expansion of the stellar 
potential in basis functions, i.e.\ essentially collisionless. Their code thus
follows the dynamical evolution of the \bh s and stars in a self-consistent way.
In their simulations, Quinlan \& Hernquist started with the stars being in 
spherical isotropic equilibrium following a Jaffe model (which has $\gamma=2$) 
and various choices for the masses and initial positions of the \bh s. Their
general conclusions relevant here are as follows. 
(i) The total mass of the stars ejected from the centre by three-body 
interactions is about twice the mass of the \bh\ binary.
(ii) Wandering of the \bh\ pair significantly increases the loss cone and
mitigates the problem of its depletion.
(iii) Inside about $r_h$, the density profile has become much shallower,
almost flat, the velocity dispersion increases in a Keplerian fashion, and
the stellar motions are significantly tangentially anisotropic with $\beta
\approx-1$ (see Fig.~\ref{quinlan_hernquist}). 

Thus, at least qualitatively, this process can explain the presence of a 
shallow stellar cusp with significant tangential anisotropy around a central
black hole. However, there are some points where M87 does not fit smoothly
into this picture. For example, the radius inside which the density of M87 
becomes shallow ($\approx10\arcsec$) is clearly larger than the radius inside 
which $\beta<0$ ($\approx3\arcsec$), while the simulations indicate that these 
radii should be similar. Also, the central cusp of M87 is not completely dug
out as in Quinlan \& Hernquist's simulations. These discrepancies indicate
that reality is more complicated, possibly involving several such accretion
events with ``small'' secondary \bh s and/or dissipational processes (e.g.\
star formation) as indicated by the presence of the central gas-disk. 

\section{Influence of a Central Black Hole on Larger Scales}
\index{black hole!interaction with host}
\index{black hole!mass limit}
The processes discussed in the last section influence the structure and 
dynamics of the stellar system only in the immediate neighbourhood of the \bh, 
i.e.\ inside $r_h$. A central \bh, however, will influence all those stars that 
ever come near the centre. Many stars in triaxial galaxies are on box orbits,
which pass arbitrarily close to the centre after a sufficiently long time. It
was argued by Gerhard \& Binney (1985) that scattering and re-distribution of
box orbits by a central \bh\ would cause at least the inner parts of a triaxial
galaxy to become rounder or more axisymmetric. 

\begin{figure}
	\centerline{ \epsfxsize=8cm \epsfbox{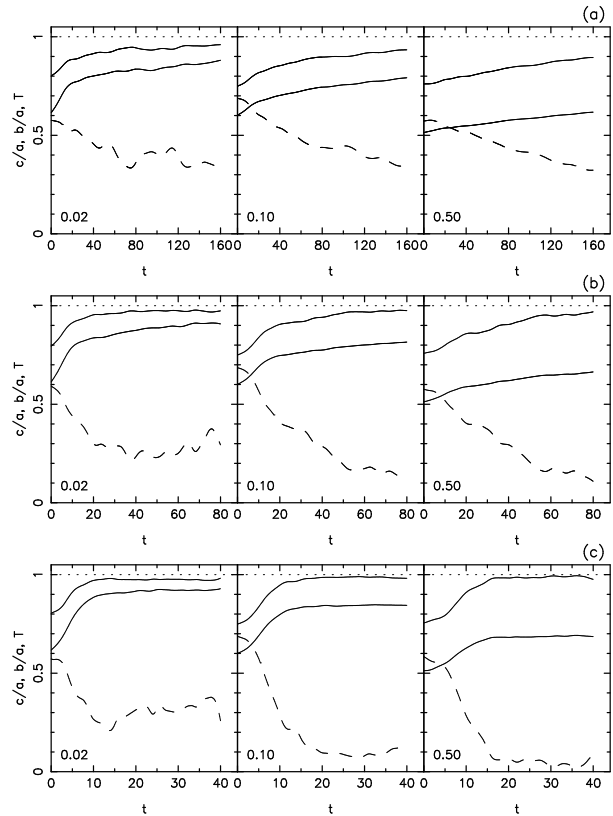} }
\caption[]{Time-evolution of the axis ratios $b/a$ and $c/a$ (full lines) and 
the triaxiality index $T=(a^2-b^2)/(a^2-c^2)$ (dashed) in the simulations of
Merritt \& Quinlan (1998). The mass ${\rm M}_\bullet$ of the \bh\ grown in 
$t_{\rm grow}=15$ is 0.003 (a), 0.01 (b), and 0.03 (c) of the total mass in 
stars. The numbers in the left corner of each frame are the fraction of 
particles, ranked by binding energy, that were used in computing the shape
parameters.} \label{merritt}
\end{figure}

Using a modification of the $N$-body code employed by Quinlan \& Hernquist,
Merritt \& Quinlan (1998) have recently studied this process numerically. 
They created a stable triaxial equilibrium model by the simulated collapse of a 
non-equilibrium configuration, and followed the time-evolution of this model 
when at its centre a point mass was slowly grown.
Fig.~\ref{merritt} shows, for three different \bh\ masses, the time-evolution 
of the shape of the model.
In all simulations, the galaxy model tends to become axisymmetric, even at the 
half mass radius. For the lightest \bh\ with ${\rm M}_\bullet/{\rm M}_{\rm g}
= 0.003$ (${\rm M}_{\rm g}$ denotes the mass of the galaxy), this process is
still ongoing at the end of the simulation, whereas for ${\rm M}_\bullet/{\rm 
M}_{\rm g} = 0.01$ it is nearly finished. The most interesting result, however,
is that for ${\rm M}_\bullet/{\rm M}_{\rm g} = 0.03$ the process is almost
finished at the time $t_{\rm grow}$ when the \bh\ has reached its final mass.
Further simulation of the authors with varying $t_{\rm grow}$ confirmed this 
result: for ${\rm M}_\bullet/{\rm M}_{\rm g}= 0.03$ the shape becomes 
axisymmetric as soon as possible, i.e.\ at $t=\max\{t_{ \rm grow},t_{\rm 
dyn}(r)\}$, where $t_{\rm dyn}(r)$ denotes the dynamical or crossing time at 
radius $r$.

This behaviour can be explained by the \bh\ inducing chaos in the orbital 
motions of the triaxial galaxy. For small ${\rm M}_\bullet$, the orbits become 
weakly chaotic, i.e.\ they behave like a box orbit over some time, but after a 
sufficiently long time they fill their energy surfaces. For ever larger ${\rm 
M}_\bullet$, the orbits become ever more strongly chaotic until at some
critical ${\rm M}_\bullet$ the Liapunov time (the time in which two 
neighbouring trajectories diverge) equals the dynamical time, i.e.\ the orbits
no longer resemble box orbits at all. In the simulations of Merritt \&
Quinlan this critical ${\rm M}_\bullet$ is about 2.5\% of ${\rm M}_{\rm g}$. 
In general, this number should depend on details of the triaxial configuration,
but is likely to be of the same order, i.e.\ $\sim10^{-2}{\rm M}_{\rm g}$.

From the existence of this critical ${\rm M}_\bullet$, Merritt \& Quinlan draw 
a very interesting conclusion. In order for \bh s to grow by gas accretion (the 
standard model for AGNs), the gas has to reach the \bh\ from large radii, i.e.\
it has to lose its angular momentum. In axisymmetric galaxies, the conservation
of angular momentum along ballistic orbits renders gas-fueling of the centre
very difficult. Thus, a \bh\ may cut off its own gas supply by changing the
shape of its host galaxy. If this picture actually applies, the \bh\ mass should
be no larger than the critical ${\rm M}_\bullet \sim10^{-2}{\rm M}_{\rm g}$.
A correlation in this sense is indeed observed: ${\rm M}_\bullet$ inferred from
the dynamics of several early-type galaxies is always of this order (cf.\
Kormendy \& Richstone 1995). In M87, ${\rm M}_\bullet/{\rm M}_{\rm g}$ is only
$\sim0.5\%$, which might be a consequence of mergers that convert disks into
spheroids and hence increasing ${\rm M}_{\rm g}$ (Merritt, private 
communications).

Several \bh s, however, are hosted by barred spiral galaxies (e.g.\ the Galaxy,
NGC 1068). Tumbling bars are mainly made of stars on so-called $x_1$ orbits,
which avoid the very centre. Hence, the mechanism working on box orbits for 
triaxial bulges may not (or not as well) work for barred spirals.

\section{Summary}
The kinematics of M87 are well studied, which make this galaxy a good test case
for the theories of galaxy formation. Outside $\sim100\arcsec$, the 
velocity dispersion profile rises indicating the presence of large amounts of
non-luminous matter. The inferred density profile $\rho\propto r^{-1.3}$ is
consistent with predictions from CDM cosmogony for the inner parts of 
dark-matter halos.

The massive black hole (\bh), detected in the very centre of
M87 by gas motions, together with the observed stellar kinematics implies a
significant tangential anisotropy of the stellar motions. Among the formation 
histories discussed for a \bh\ in a galactic centre, only the model of accretion
of other massive \bh s, originating from the centres of cannabalized companions,
can explain such a strong anisotropy. This scenario also predicts a shallow 
stellar density cusp as observed for M87. (Quantitatively, there are 
some discrepancies, which may well be due to over-simplification in the
simulations of this process.)

A massive \bh\ at the centre of a triaxial galaxy renders, by the destruction of
box orbits, the shape of its host axisymmetric. This mechanism becomes very
fast once the \bh\ mass reaches a critical value, which is of the order of 1\% 
of its host's mass. Since the conservation of angular momentum along ballistic
orbits in an axisymmetric galaxy obstructs gas-fueling of the centre, this
process may pose an upper limit for the mass a \bh\ can reach by gas-accretion.
An upper limit of this order is indeed observed among \bh\ masses inferred from
the dynamics of early-type galaxies.

\section*{Acknowledgements}
I am grateful to the organizers for inviting me to this wonderful workshop.
Special thanks to David Merritt, who helped improving on an early version
and made Figs.~3 and 5 available in electronic format.

\end{document}